\documentclass[12pt]{article}
\usepackage{epsfig}
\newcommand{\beq}{\begin{equation}}
\newcommand{\eeq}{\end{equation}}

\newcommand{\be}{\begin{eqnarray}}
\newcommand{\ee}{\end{eqnarray}}
\newcommand{\Lagr}{{\mathcal L}}
\newcommand{\dd}{\mathrm d}

\begin{document}
\title{\bf Baryonic masses based on the NJL model}
\author{W.M. Alberico, 
F. Giacosa\footnote{Present address: Institut f\"ur Theoretische Physik,
Universit\"at T\"ubingen, Auf der Morgenstelle 14, 72076 T\"ubingen, Germany}, 
M. Nardi,  C. Ratti\\
\vspace*{1cm}\\
%\affiliation
{\it Dipartimento di Fisica Teorica, Universit\`a di Torino}\\
{\it and INFN, Sezione di Torino, 
Via P. Giuria 1, I-10125 Torino, Italy}\\
}

\date{\today}
\maketitle

\begin{abstract}
We employ the Nambu Jona--Lasinio model to determine the vacuum pressure
on the quarks in a baryon and hence their density inside. Then we estimate
the baryonic masses by implementing the local density approximation for 
the mean field quark energies obtained in a uniform and isotropic system.
We obtain a fair agreement with  the experimental masses.
\end{abstract}

\vspace*{1cm}
{\bf PACS:} 12.39.-x, 12.39.Ba, 12.39.Fe, 14.20.-c, 14.20.Dh\\

\section{Introduction}
The Nambu Jona--Lasinio (NJL) model  is a phenomenological quark model, 
which entails chiral symmetry breaking at low density and temperature, 
and chiral symmetry restoration at high density and  
temperature. In the chirally broken phase, quarks develop a dynamical mass 
by their interaction with the vacuum.\cite{NJL,Buballa98}

This model has been employed in many different contexts, for calculations of 
meson properties, for hot and dense matter and for the study of 
diquarks (for reviews about the model and its applications see, for example,
\cite{Klev92,Vogl91,Hatsuda94}). 
It can successfully reproduce various empirical aspects
of QCD such as the non--perturbative vacuum structure, dynamical breaking of 
chiral symmetry, the $U_A(1)$ anomaly and explicit flavour $SU(3)$ breaking in
the hadronic spectrum. The strong attractive force between quarks in the 
$J^P=0^+$ channel induces an instability in the Fock vacuum of massless 
quarks, thus generating a dynamical mass $m^*$ which is typically of the
order of a few hundred MeV, in agreement with the constituent quark 
masses~\cite{Shuryak84,Pagels75,Christ84,thoof86,Gasser82}.

This latter feature allows one to consider the constituent quark basis 
generated by dynamical symmetry breaking as a good starting point for the
description of hadrons, including baryons. However, beyond the conventional
constituent quark model, the NJL model  takes into account the collective
nature of the vacuum and the Nambu-Goldstone bosons. The remaining 
non--perturbative effects such as the long--range confinement force can be
treated as a relatively weak perturbation for low-energy phenomena.
A similar picture has been first applied to baryons in Ref.~\cite{Kuni90}.

 In this work, we first connect the NJL model to the evaluation
of the typical dimensions of a baryon (in particular the nucleon radius). 
Then, in the spirit of the constituent quark model, we employ this result 
to calculate the octet baryon masses.

In Section 2 we recall the main results of the NJL model, such as the 
chiral condensate for quarks, the expression for the dynamical masses, the 
energy and pressure of a uniform and isotropic quark system. As it is known, 
the dynamical masses in the vacuum are quite large compared to the current
quark masses (for the quarks $u$ and $d$ one goes from a current quark mass
of 5~MeV to an effective mass larger than 300~MeV), but when the density 
increases the dynamical masses decrease and hence the degree of chiral 
symmetry  breaking, caused by the large effective masses, becomes less severe.
A complete chiral restoration can only be achieved when the bare masses $m_i$
vanish, but even very small values of $m_i$ are enough to prevent it.

We want to apply these considerations to the problem of the stability of the 
nucleon. First we note that outside the nucleon there is a vacuum pressure 
($P_{vac}$) caused by the negative energy of the Dirac sea, but inside the 
nucleon the pressure due to the Dirac sea ($P_{N,vac}$) is much lower since it 
is a high density region and the dynamical masses become small. As a 
consequence one has an effective pressure acting on the nucleon, due to the 
difference between  the energy densities.

The situation is similar to the one encountered in the MIT bag 
model~\cite{Fahri84,Chodos74,Chodos74b,Schertle99}, but here the
bag pressure is by itself a model parameter, while 
in the NJL model the pressures ($P_{vac}$, $P_{N,vac}$) are deduced after 
tuning the other parameters to reproduce, e.g., the experimental 
mesonic masses. For a generic attempt to derive the bag pressure see 
Ref.~\cite{Hof00}.

The residual effective pressure is equilibrated by the pressure generated by 
the three quarks inside the nucleon. Obviously one can not consider them as a 
uniform and isotropic quark system, neither express their
pressure with the formulas developed in Section 2. For this purpose, in 
Section 3 we will assume heuristic wave functions for the quarks confined 
in the nucleon.
Under this assumption we can calculate the quark pressure and, by imposing 
that it equals the effective vacuum pressure, we can derive the radius of the 
nucleon, to be compared with the experimental value. 

In Section 4 we develop the calculation for the masses of the baryonic 
octet, utilizing the results of Section 3 and  
taking into account the dynamical quark masses. The latter are derived by 
implementing the local density approximation on a uniform gas of quarks to
obtain the dependence of the masses upon the distance from
the center of the baryon, according to the spatial quark density found in 
the previous Section. The masses increase with the distance, since the density
decreases and vanishes outside the baryon. An analogous procedure is utilized 
for the kinetic energy contribution to the baryonic mass. 

 We do not attempt, here, an approach based on the relativistic Faddeev 
model, quite often implemented from a NJL-type Lagrangian and applied both 
to the nucleon~\cite{Ishii93,Ishii95,Asami95} and to the baryons' octet and
decuplet~\cite{Krew95}. The merit of the Faddeev approach is to offer an 
evaluation of the quark wave function which is consistent with the model 
Lagrangian employed. Mesonic properties and the masses of the nucleon and 
$\Delta$ resonance are well reproduced. Yet, the constituent quark 
masses are kept constant for a given parameter set. 

The central and key point of the present work, instead, is to take into 
account the variation of the constituent masses with the density of quarks
inside the hadron: independently of the details of the wave functions 
employed here, this appears to be crucial in order to get a realistic 
determination of the baryonic masses. Some warning should apply to the
use of the NJL at finite densities (as it has been extensively discussed in 
ref.~\cite{Klev92}), where it can lead to unphysical results, like 
having, e.g., vanishing or even negative effective mass of the quark.
We shall shortly discuss this point in connection with the present 
calculation.

Finally we also calculate 
the masses of two baryons from the spin ${3}/{2}$ decuplet, and find that the
present calculation provides for them a reasonable estimate, which can be 
slightly improved by the effect of the spin interaction.
We summarize and discuss our results in Section 5.

\section{Short review of the NJL model}
Many versions of the NJL Lagrangian have been used in the past, particularly 
in the last ten years; we consider a three flavour NJL Lagrangian of the 
form~\cite{Klev92,Vogl91,Hatsuda94,Hufner96}:
\beq
{\Lagr}_{NJL}={\Lagr}_0+{\Lagr}_m+{\Lagr}_{(4)}+{\Lagr}_{(6)}
\label{Lagreq}
\eeq
where:
\be
\Lagr_0 &=& i \,\bar{q}\, \gamma^\mu\,\partial_\mu q\, ,
\label{L0}\\
\Lagr_m &=& - \,\bar{q}\,{\bf\hat{m}}\,q\,,
\label{Lm}\\
\Lagr_{(4)}&=& \frac{G}{2} \sum\limits_{k=0}^8 \,
\left[\left(\bar{q}_i(\lambda_k)_{ij}q_j\right)^2+
\left(i\bar{q}_i\gamma_5(\lambda_k)_{ij}q_j\right)^2\right]\,,
\label{L4}\\
\Lagr_{(6)}&=& -K\left[\det_{i,j}\left(\bar{q}_i(1+\gamma_5)q_j\right)+
\det_{i,j}\left(\bar{q}_i(1-\gamma_5)q_j\right)\right]\,.
\label{L6}
\ee
In the above
$q \equiv \left( {{\begin{array}{c} u \\ d \\ s\end{array}}} \right)$ 
is the quark field,
${\bf\hat{m}}    \equiv\left({{\begin{array}{ccc}
m_u & 0 & 0 \\0&m_d&0\\0&0&m_s\end{array}}}\right)$ is the mass matrix,
$\lambda_1\ldots\lambda_8$ are the Gell--Mann flavour matrices,
and $\lambda_0\equiv\sqrt{\frac{2}{3}}{\mathbf 1}$.

$\Lagr_{(4)}$ generates four--leg interaction vertexes, while 
$\Lagr_{(6)}$ gives rise to six--leg interaction vertexes; $G$ and $K$ 
are two parameters of the model, with the dimensions of $[L^{2}]$ and 
$[L^{5}]$, respectively.
In the limit of $m_i=0$, the symmetries of the model Lagrangian are the 
following ones:
\be
U_V(1)\times SU_V(3)\times SU_A(3)\times SU_c(3)
\ee
where, of course, $SU_c(3)$ is global and not local; $U_A(1)$ is broken by
the existence of the axial anomaly. For an extended analysis of this NJL 
Lagrangian, see for example Ref.~\cite{Buballa98}.

Within the mean field approximation it is possible to evaluate the dynamical 
quark masses, which are generated by the quark--vacuum interaction;
 we consider homogeneous, isotropic quark matter, with Fermi 
momenta $p_F^u, p_F^d$ and $p_F^s$ for the flavours $u, d$ and $s$, 
respectively. The result of the calculation is~\cite{Schertle99}
($i, j, k = u, d, s$):
\be
m_i^* = m_i-2G\langle\bar{q}_i q_i\rangle+2 K 
\langle\bar{q}_j q_j\rangle\langle\bar{q}_k q_k\rangle
\qquad\quad (i\ne j\ne k)
\label{mstari}
\ee
where
\be
\langle\bar{q}_i q_i\rangle = -\frac{3}{\pi^2}
\int\limits_{p_F^i}^\Lambda
\dd p \frac{m_i^* p^2}{\sqrt{p^2+(m_i^*)^2}}
\label{condens}
\ee
 is the chiral condensate for the $i$-flavour.
We introduce a three--dimensional regularization with a cut--off $\Lambda$
since the integral is obviously divergent. 
It is important to note that as the Fermi momenta increase, the chiral 
condensate and the dynamical masses decrease. In the vacuum the dynamical 
masses reach their highest values.

For our purposes it is very important to consider the energy density of 
the system, which is given by the following formula~\cite{Schertle99}:
\be
\varepsilon=-\sum\limits_{i=u,d,s}\frac{3}{\pi^2}\int\limits_{p_F^i}^\Lambda
\dd p p^2\sqrt{p^2+(m_i^*)^2} +
\left[\sum\limits_{i=u,d,s} \,G
\langle\bar{q}_iq_i\rangle^2\right]
-4K\langle\bar{u}u\rangle \langle\bar{d}d\rangle \langle\bar{s}s\rangle.
\label{enden}
\ee

The pressure of the system can then be derived to be:
\be
P=-\varepsilon+\rho_i\mu_i
\label{press}
\ee
where:
\be
\rho_i=\frac{\left(p_F^i\right)^3}{\pi^2}
\ee
is the quark density of flavour $i$ and 
$\mu_i=\sqrt{\left(p_F^i\right)^2+(m_i^*)^2}$ the corresponding chemical 
potential. The pressure of vacuum is 
then, using formula (\ref{enden}):\footnote{Several 
works\cite{Buballa98,Schertle99} 
introduce a constant, $B$, which allows to set the vacuum 
pressure and energy to zero; this procedure, however, is not useful for our 
purposes, since we are mainly interested in differences of pressures, 
and we will not make use of it.}
\be
P_{vac}=-\varepsilon_{vac}= - \varepsilon(p^i_F=0),
\ee
which is positive for the customary choices of the parameters $G$ and $K$.

In the following Sections we will consider two sets of parameters, 
which have been employed in Refs.~\cite{Hufner96} (set 1) 
and \cite{Hatsuda94} (set 2); in both cases the current masses for the 
$u$ and $d$ quarks 
are fixed on the basis of isospin symmetry and of limits on the average
$\bar m=(m_u+m_d)/2$ at 1~GeV scale, while the remaining four parameters
are fitted to reproduce the masses of $\pi$,
$K$ and $\eta'$ mesons, together with the pion--decay constant 
$f_\pi$:\footnote{Notice that in other works the NJL Lagrangian is
written using different notations (e.g.
$G$ instead of $G/2$ or with different sign for the six--quark coupling).
Obviously this must be taken into account in comparing the numerical 
values of the parameters.}
\begin{center}
\begin{tabular}{|c|c|}
\hline
{\bf set 1} &{\bf set 2}\\
\hline
$m\equiv m_u=m_d=5.5$ MeV & $m\equiv m_u=m_d=5.5$ MeV \\
$m_s=140.7$ MeV           & $m_s=135.7$ MeV\\
$\Lambda=602.3$ MeV          & $\Lambda=631.4$ MeV\\
$G\Lambda^2=3.67$        & $G\Lambda^2=3.67$ \\
$K\Lambda^5=12.36$        & $K\Lambda^5=9.29$\\
\hline
\end{tabular}
\end{center}

With the above values, the effective quark masses in the vacuum and the chiral
condensates turn out to be:

\begin{center}
\begin{tabular}{|c|c|c|}
\hline
&{\bf set 1}(MeV) &{\bf set 2}(MeV)\\
\hline
$m^*_u=m^*_d$ & 367.7 & 335.5\\
$m^*_s$ & 549.5 & 527.6\\
$|\langle{\bar q}q\rangle_u|^{1/3}=|\langle{\bar q}q\rangle_d|^{1/3}$ & 241.9 
& 246.7\\
$|\langle{\bar q}q\rangle_s|^{1/3}$ & 257.7 & 266.7\\
\hline
\end{tabular}
\end{center}

\section{Nucleon radius}
We will heuristically assume that a quark confined in a nucleon has a 
Gaussian wave function; the reasons for this assumption are essentially three: 
this wave function reproduces the confinement of a particle in a spatial 
region, it can be treated a\-na\-ly\-ti\-cally, and it is also the ground 
state wave function of an harmonic oscillator. Hence, with the correct 
normalization, the spatial wave function of the three valence quarks in a 
nucleon is written as:
\be
\Psi_q\left(r\right)=
\left(\frac{2}{\pi}\right)^{\frac{3}{4}}\frac{1}{b^{\frac{3}{2}}}
e^{-\frac{r^2}{b^2}}
\label{psiq}
\ee
where $b$ is a parameter with the dimension of a length.
The total baryonic density is then:
\beq
\rho_B(r)=3\left(\frac{1}{3}\right)\left|\psi_q(r)\right|^2
=\left|\psi_q(r)\right|^2,
\eeq
which coincides with the probability density for one quark. 
To fix the parameter $b$ we establish a connection with the properties
derived in the NJL model. 

First  we calculate the following average 
quantities (average squared radius and average volume)
from $\Psi_q\left(r\right)$:
\be
\langle r^2 \rangle=\frac{3}{4}b^2; \hspace{1cm}
\langle V\rangle=\frac{4}{3}\pi\langle r^2\rangle^{3/2}=
\frac{\pi\sqrt{3}}{2}b^3.
\label{averages}
\ee

Then, by considering the Fourier transform of $\Psi_q\left(r\right)$,
\be
A(k)=\frac{b^{3/2}}{(2\pi)^{3/4}}e^{-\frac{1}{4}k^2b^2}
\ee
we can calculate the average  kinetic energy of the quark as follows:
\be
\langle E_q\rangle=4\pi \int\limits_0^\infty \dd k
k^2\sqrt{(m^*_q)^2+k^2} \left|A(k)\right|^2
\ee
where $m^*_q$ is the dynamical quark mass. 

As we have already noticed in the previous Section, the dynamical mass of 
a quark in a high density region is small, and this is precisely the 
situation of the quark inside a baryon; therefore, we can eventually 
neglect $m^*_q$ in the previous formula, and explicitly obtain, for 
the average energy of the quark, the analytic expression:
\be
\langle E_q\rangle \simeq 4\pi 
\int\limits_0^\infty \dd k\left|A(k)\right|^2  k^3 = 
\frac{4}{\sqrt{2\pi}}\frac{1}{b}= \frac{\gamma}{\langle V \rangle^{1/3}}
\label{Equark}
\ee
where use has been made of relation (\ref{averages}) to express 
$\langle E_q\rangle$ in terms of the average volume and
$\gamma=(4/\sqrt{2\pi})\left(\pi\sqrt{3}/2\right)^{1/3}$.
 In this approximation
the total energy density of a nucleon turns then out to be:
\be
\langle E_T\rangle = 3\langle E_q\rangle,
\ee
from which we can calculate the pressure of the three--quark system 
as follows:

\be
P=-\left(\frac{\partial\langle E_T\rangle}{\partial \langle V \rangle}
\right)_{N_q} =\frac{\gamma}{\langle V \rangle^{4/3}}.
\label{pressure}
\ee
$N_q$ being the total (fixed) number of quarks. 
This quantity obviously depends on the parameter $b$.

In order to set a connection with the NJL model, we shall now evaluate
 the effective vacuum pressure acting on the nucleon. 
In the NJL model  the energy density can be expressed 
in the following way:
\be
\varepsilon & =& \varepsilon_{u,d}+\varepsilon_s 
\label{enerden}\\
\varepsilon_{u,d}  & =&- \sum\limits_{i=u,d} \frac{3}{\pi^2}
\int\limits_{p_F^i}^\Lambda \dd p \,
p^2\sqrt{p^2+(m_i^*)^2} + \sum\limits_{i=u,d}
G\langle\bar{q}_iq_i\rangle^2-
2K\langle\bar{u}u\rangle \langle\bar{d}d\rangle \langle\bar{s}s\rangle
\nonumber\\
\varepsilon_s& =& -\frac{3}{\pi^2}\int\limits_0^\Lambda \dd p \,
p^2\sqrt{p^2+(m_s^*)^2}
+G\langle\bar{s}s\rangle^2-\langle\bar{s}s\rangle\left(
m_s^*-m_s+2G\langle\bar{s}s\rangle\right)
\nonumber
\ee
where we have set $p_F^s$=0, since in the nucleon
no strange valence  quarks are present.
It is possible to show~\cite{Buballa99} that in this case 
$\langle\bar{s}s\rangle$ is almost constant with varying $u$ and $d$ 
densities. In turn, this implies:
\beq
\frac{\dd\varepsilon_s}{\dd m_{s}^{*}}=
\langle\bar{s}s\rangle-\langle\bar{s}s\rangle=0.
\eeq
By considering that $m^*_s$ depends on $\rho_u$ and $\rho_d$ [see 
eqs.~(\ref{mstari}), (\ref{condens})], it follows that:
\beq
\frac{\partial\varepsilon_s}{\partial\rho_u}=
\frac{\partial\varepsilon_s}{\partial\rho_d}=0.
\eeq
For this reason $\varepsilon_s$ does not contribute to the pressure acting 
on the nucleon (it has the same value inside and outside)
and in the following  we will consider only the
$\varepsilon_{u,d}$ contributions. 

In the vacuum, with $\rho_u=\rho_d=0$ (and $p_F^u=p_F^d=0$), the $u,d$ 
contribution to the pressure is then:
\be
P_{vac}=-\varepsilon_{u,d(vac)}
\ee
which, for the above mentioned two different sets of parameters, turns 
out to be:
\beq
\begin{array}{l}
P_{vac}  =  2.091\times 10^{10}\,\,({\mathrm {MeV}})^4
\hspace{2cm}{\mbox{(set 1)}}\\
P_{vac}  =  2.493\times 10^{10}\,\,({\mathrm {MeV}})^4
\hspace{2cm}{\mbox{(set 2)}}
\end{array}
\label{Pvac}
\eeq

Inside the nucleon the vacuum pressure is uniquely related to the Dirac 
sea energy density; indeed, since in the nucleon interior $u$ and $d$ 
quark densities are high, we can neglect, as before, their 
chiral condensate and their masses. Hence we have the following expression 
for the internal pressure:
\be
P_{N,vac}=2\frac{3}{\pi^2}\int\limits_0^\Lambda \dd p\,p^3=
\frac{3}{2\pi^2}\Lambda^4
\label{Pressvac}
\ee
with the following numerical values:
\beq
\begin{array}{l}
P_{N,vac}  =  2.00\times 10^{10}\,\,({\mathrm {MeV}})^4
\hspace{2cm}{\mbox{(set 1)}}\\
P_{N,vac}  =  2.42\times 10^{10}\,\,({\mathrm {MeV}})^4
\hspace{2cm}{\mbox{(set 2)}}.
\end{array}
\label{PNvac}
\eeq
Finally the effective vacuum pressure acting on the nucleon is:
\be
P_{eff}=P_{vac}-P_{N,vac}
\ee
for which we get: 
\beq
\begin{array}{l}
P_{eff}  =  9.13\times 10^{8}\,\,({\mathrm {MeV}})^4
\hspace{2cm}{\mbox{(set 1)}}\\
P_{eff}  =  7.71\times 10^{8}\,\,({\mathrm {MeV}})^4
\hspace{2cm}{\mbox{(set 2)}}.
\end{array}
\label{Peff}
\eeq
 These values can be compared to the ones of the $B$ parameter in the 
MIT bag model: the authors of Ref.~\cite{degrand75} employed a value 
$B=4.42\times 10^8$~MeV$^4$ for reproducing masses and other parameters
of light hadrons, while the value $B=7.68\times 10^8$~MeV$^4$ (very close 
to our net pressure with parameter set 2) was more
appropriate for the hadronic structure functions~\cite{Steffens95}.

In order to have equilibrium, the effective vacuum pressure and the 
three quark pressure (\ref{pressure}) must be equal; by imposing this 
condition we can then fix the parameter $b$ in terms of the NJL model
parameters. We obtain the following values:
\be
b&=&0.993\,\,{\mathrm {fm}}\hspace{1cm}{\mathrm{(set 1)}}
\nonumber\\
b&=&1.036\,\,{\mathrm {fm}}\hspace{1cm}{\mathrm{(set 2)}}.
\nonumber
\ee

We can calculate $\langle r^2 \rangle$ and hence the radius 
of the nucleon, $R=\sqrt{\langle r^2\rangle}$, which turns out to be, 
for the two parameter sets:
\be
\nonumber
\sqrt{\langle r^2 \rangle} & = & 0.860~~ \mathrm{fm~~~~~~~~~set} 1\\
\nonumber
\sqrt{\langle r^2 \rangle} & = & 0.897~~ \mathrm{fm~~~~~~~~~set} 2
\ee
in fair agreement with the experimental determination, 
$\sqrt{\langle r^2\rangle_{exp}}$=0.81 fm.

In spite of the simplicity of our approach, this result points to an
interesting interpretation, since both the Dirac sea and the vacuum pressure 
seem to play an important role in the nucleon stability.

In the next section we will use this model
for the calculation of the baryon octet masses. 
The above determination of the value of $b$, in particular the approximation
 of neglecting the quark masses inside the nucleon, employed in 
eqs.~(\ref{Equark}) and (\ref{Pressvac}),
will be discussed: indeed it poses a delicate
problem of self--consistency, as it will be clarified below.

\section{Octet baryon masses}

In order to evaluate the baryon masses we sum the energies of the
constituent quarks: for this purpose we recall that in the NJL model the 
effective mass, at fixed density, is given by eqs.~(\ref{mstari}), 
(\ref{condens}). We shall implement these expressions by taking into account 
the quark density inside the baryon, according to the Ansatz (\ref{psiq}) for 
the quark wave function.

The density of $N_i$ quarks of flavour $i$ is then:
\be
\rho_i(r)=N_i\left|\psi_q^{i}(r)\right|^2,
\label{density}
\ee
with the condition: $N_u+N_d+N_s=3\,$.
We can  obtain the behaviour of the dynamical mass $m_i^*(r)$ 
as a function of the distance $r$ from the center of the baryon by using the
so--called Local Density Approximation (LDA). It amounts to define a 
local Fermi momentum $p_F^i(r)=[\pi^2\rho_i(r)]^{{1}/{3}}$ for each flavour
of quarks and then insert it into the self--consistent definition of the
dynamical mass, which thus becomes $(i\ne j\ne k)$:
\be
m_i^*(r)&&= m_i + \frac{6G}{\pi^2}
\int\limits_{p_F^i(r)}^\Lambda \dd p \frac{m_i^*(r)p^2}
{\sqrt{p^2+(m_i^*(r))^2}} + 
\label{mstarloc} \\
&&+ 2 K \left(\frac{3}{\pi^2}\right)^2
\left[\, \int\limits_{p_F^j(r)}^\Lambda \dd p \frac{m_j^*(r)p^2}
{\sqrt{p^2+(m_j^*(r))^2}}\right]
\left[\, \int\limits_{p_F^k(r)}^\Lambda \dd p \frac{m_k^*(r)p^2}
{\sqrt{p^2+(m_k^*(r))^2}}\right]
;
\nonumber
\ee
the $r$ dependence of $m_i^*$ is governed by the lower limits of 
integration and by the self-consistency requirement entailed by 
eq.~(\ref{mstarloc}) itself.
\begin{figure}[htb]
\begin{center}
\mbox{\epsfig{file=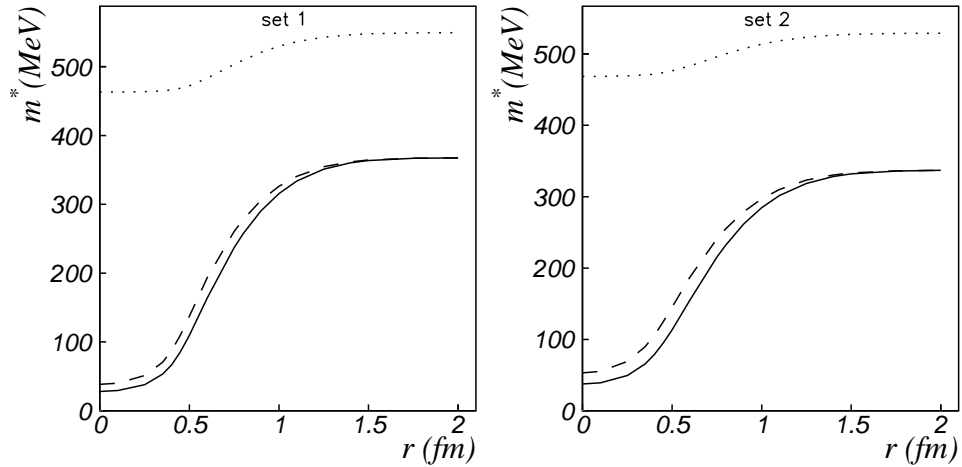,width=0.85\textwidth}}
\end{center}
\caption{Quark masses as a function of $r$ in the nucleon case for the two 
different sets of parameters. In both panels, solid lines represent $m_u$, 
dashed lines $m_d$ and dotted lines $m_s$. }
\label{fig1}
\end{figure}

As an example, we show in Fig.~\ref{fig1} the masses of the $u$, $d$ and $s$
quarks in the proton ($N_u=2$, $N_d=1$, and $N_s=0$) as a function of $r$, 
using the $b$ values obtained in the previous section.
One can see that the light quark masses are small for $r\sim 0$, which roughly
justifies the approximations of the previous Section. The $s$ quark mass is 
affected by the quark density in the nucleon only by the six--quark 
coupling term in the Lagrangian, hence it shows a rather mild variation from 
the interior to the exterior of the nucleon. On the contrary, the $u$ and 
$d$ masses are much smaller inside  the baryon than outside it.

The density dependence of the effective quark masses, which here and in the
following turns out to be one of the most relevant features, deserves some 
words of criticism. Indeed in the NJL model, the interplay between chemical 
potential (which defines the Fermi energy and contains the effective mass) 
and baryon density crucially depends not only on the order of 
approximation made in solving, for example, the gap equation, but also on the 
specific values of the model parameters employed. This point has been widely
and clearly discussed in the review of Klevansky~\cite{Klev92}, where the
qualitatively different outcomes (e.g. for the quark effective mass
as a function of the chemical potential) are elucidated. In particular it is 
shown that: {\it i)} the order of the phase transition which occurs as the 
density increases depends upon the inclusion of exchange effects in the mean 
field approximation and {\it ii)} the properties of the system with increasing 
density dramatically depend upon the chosen set of parameters. Of special
relevance is the value of the cutoff $\Lambda$ as compared with the 
chemical potential, since with increasing density a too small cutoff could
lead to unphysical negative effective masses. These problems do not 
specifically affect the results of the present work, but were carefully
considered and checked at any stage of the calculation.

To calculate baryonic masses we still need the ``local'' kinetic 
energy for the quark; in order to avoid an additional integration over 
the Fermi sphere for 
each quark species, we employ the average momentum in the local Fermi sphere,
namely $(3/5){p_F^i}^2$. Hence the total baryonic mass reads:
\beq
M=\sum_{i=u,d,s}\int\limits_0^\infty \dd r~4\pi r^2\rho_i\left(r\right)
\sqrt{\frac{3}{5}p_F^i\left(r\right)^2+(m_i^*\left(r\right))^2}.
\label{Bmass}
\eeq
\begin{table}[t]
\begin{center}
\begin{tabular}{c|cccccc}
\hline
\hline
Baryon & N &$\Lambda$ &$\Sigma^{+}$ &$\Delta^{++}$ &$\Xi^{0}$ &$\Omega^{-}$ \\
\hline \\
$m_{exp}$~(MeV) & 938.27 &1115.68 &1189.37 &1232 &1314.9 &1672.45 \\
\hline \\
$m_{set\, 1}$~(MeV) & 981.22 & 1150.59 & 1199.20 & 1072.28 & 1349.78 & 1536.84 \\
%\hline
 \\
$m_{set\, 2}$~(MeV) & 935.15 & 1119.71 & 1159.12 & 1013.36 & 1325.96 & 1517.38\\
%\hline
\hline
\hline
\end{tabular}
\end{center}
\begin{center}
\footnotesize
Table I. Experimental masses and theoretical masses of
baryons calculated in the NJL model.
\end{center}
\end{table}
With this procedure, we can perform the calculation for the whole baryonic 
octet simply by changing the (valence) quark numbers $N_i$. 
Note that the dynamical mass of 
the $u$-quark, for example, depends also on the $d$ and $s$ condensates;
therefore there is a strong interplay among the different components of 
the system. The results of our calculations are shown in the second and third
rows of Table I.
As one can see, the results for the octet masses are in very good 
agreement with the experimental values, for both sets of parameters.

One should notice, however, that the above results are based on an 
approximate procedure in evaluating the $b$ parameter for up and down quarks
in the proton, not fully consistent with the local density approximation. 
Moreover the same wave function of the $u$, $d$ quarks has been utilized for 
the $s$--quark in all considered
hyperons. After having introduced the LDA for the
quark masses, a test of consistency is in order, both to check the 
approximations employed in eqs.~(\ref{Equark}) and (\ref{Pressvac}) and to
inquire whether the $s$--quark wave function should eventually let be 
different from the $u$ and $d$--quarks one.

To test these issues, let us first consider the proton, with up and down 
quarks only: we start from eq.~(\ref{mstarloc}) by taking into account also
the $b$--dependence of the Fermi momenta (and hence of the quark masses), 
through their relation with the quark wave functions. Hence, by replacing
$p_F^i(r)\to p_F^i(r,b)$ and $m^*_i(r)\to m^*_i(r,b)$, we recalculate 
$\langle E_q\rangle$ with non--vanishing and $b$--dependent $m^*$. Then, 
following the 
same steps illustrated in section 3, we obtain a new evaluation of the 
vacuum pressure, both inside and outside the proton. We notice that here
$P_{vac}$ and $P_{N,vac}$ are (numerically) calculated by taking into account 
the non--zero values of the masses $m^*_i$ and of the condensates 
$\langle{\bar{q_i}}q_i\rangle$.

We obtain the following values for the $b$ parameter of the $u$ and $d$ 
quarks in the proton:
\be
b_{u,d}&=&0.878\,\,{\mathrm {fm}}\hspace{1cm}{\mathrm{(set 1)}}
\label{bud1}\\
b_{u,d}&=&0.922\,\,{\mathrm {fm}}\hspace{1cm}{\mathrm{(set 2)}}.
\label{bud2}
\ee
These values are typically smaller by about 11\% with respect to the ones
obtained in section 3; the corresponding proton mean square radius turns out
to be $\sqrt{\langle r^2 \rangle}=0.76$~fm (set 1) or 
$\sqrt{\langle r^2 \rangle}=0.80$~fm (set 2) still in very good agreement
with the experimental charge radius. 

Turning now to the strange quark wave function inside hyperons, we kept the
functional form (\ref{psiq}) yet allowing for a different spatial extension
of the $s$--quark distribution, namely for a different value of the parameter 
$b$. Hence, keeping the $u$, $d$ wave functions as determined in the 
nucleon, we repeated, e.g. for the $\Lambda$ hyperon, the self--consistent
determination of $b_s$ according to the above outlined procedure for the 
calculation of the internal and external pressure.
The resulting values, obtained for the two set of parameters, are:
\be
b_{s}&=&0.697\,\,{\mathrm {fm}}\hspace{1cm}{\mathrm{(set 1)}}
\label{bs1}\\
b_{s}&=&0.691\,\,{\mathrm {fm}}\hspace{1cm}{\mathrm{(set 2)}}.
\label{bs2}
\ee
They are smaller (by about $20 \div 25 \%$) of the corresponding parameter
for the up and down quarks, thus indicating a less diffuse distribution of
the strange quark inside the baryon. This outcome is quite sound, since
the strange quarks, having a larger bare (and effective) mass, are seemingly 
confined to a smaller region of space and provide a weaker kinetic pressure 
to balance the effective vacuum pressure. Analogous calculations for the
remaining hyperons in the octet as well as for the $\Omega$ baryon led us to
almost identical values for the $b$ parameter of the $s$--quark wave function,
thus indicating that the major influence comes from the mass rather than from 
the density of the strange quarks. 

\begin{table}[t]
\begin{center}
\begin{tabular}{c|cccccc}
\hline
\hline
Baryon & N &$\Lambda$ &$\Sigma^{+}$ &$\Delta^{++}$ &$\Xi^{0}$ &$\Omega^{-}$ \\
\hline \\
$m_{exp}$~(MeV) & 938.27 &1115.68 &1189.37 &1232 &1314.9 &1672.45 \\
\hline \\
$m_{set\, 1}$~(MeV) & 970.86 & 1096.34 & 1160.89 & 1095.56 & 1274.51 & 1493.10 \\
%\hline
 \\
$m_{set\, 2}$~(MeV) & 928.03 & 1067.12 & 1128.64 & 1037.89 & 1261.68 & 1486.26\\
%\hline
\hline \\
$m^{spin}_{set\, 1}$~(MeV) & 938.27 & 1063.75 & 1132.26 & 1128.15 & 1244.73 & 1522.23 \\
\hline
\hline
\end{tabular}
\end{center}
\begin{center}
\footnotesize
Table II. Experimental masses and theoretical masses of
baryons calculated in the NJL model with the parameters (\ref{bud1}), (\ref{bs1})
for the set 1, or  (\ref{bud2}), (\ref{bs2}) for the set 2.
\end{center}
\end{table}

\begin{figure}[htb]
\begin{center}
{\epsfig{file=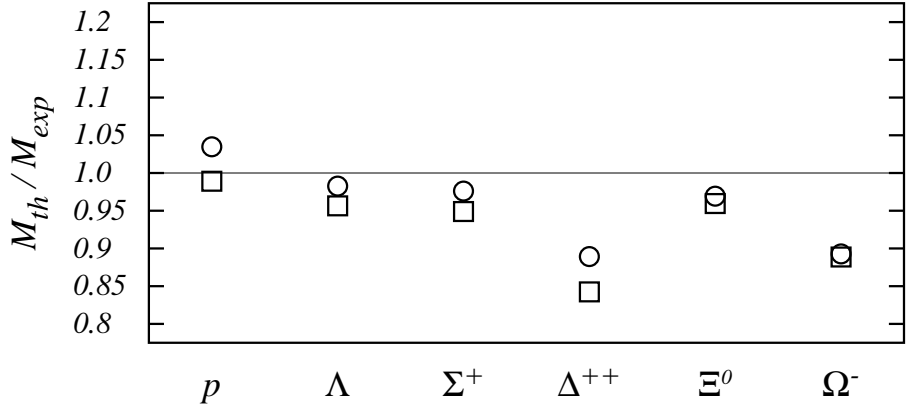,width=.85\textwidth}}
\caption{Theoretical-experimental baryon mass ratio:  
circles correspond to the first set of parameters, squares to the second 
set of parameters. }
\label{fig2}
\end{center}
\end{figure}

We have thus re-evaluated the baryonic masses according to eq.~(\ref{Bmass}),
by utilizing the parameter sets 1 and 2, together with the corresponding 
values (\ref{bud1}), (\ref{bs1}) and (\ref{bud2}), (\ref{bs2}) for the up, 
down and strange quarks, respectively, contained in the different baryons. 
The results are
reported in Table II, while in Fig.~\ref{fig2} we show the corresponding 
theoretical/experimental ratios of the baryonic masses. 
Again our results for the octet masses are in very good 
agreement with the experimental values, for both sets of parameters: by 
comparing Table I and II one can see that the self--consistent calculation 
tends to slightly decrease the theoretical masses and somewhat improves the 
agreement with the experimental ones in a few cases, among which, notably, 
the proton.

According
to Refs.~\cite{Hatsuda94,Hufner96}  we employ $m_u=m_d$, and hence we 
cannot reproduce the mass splitting between proton and neutron, or among 
baryons of other isospin multiplets.  
We also notice that a satisfactory spectrum has been obtained 
without taking into account spin corrections, which should be irrelevant
inside the same spin--parity multiplet. Finally 
we like to stress that the results presented here do not
appreciably depend upon the Ansatz for the functional form of the 
quark wave-function: indeed
calculations performed with exponential wave-functions differ only
by few percent from the ones shown here. 

As a straightforward extension of the present approach, we
have also evaluated the $\Delta$ and $\Omega$ masses, which belong to the
spin-$\frac{3}{2}$ decuplet. The results are indicated, together with the
spin $\frac{1}{2}$ octet, in Fig. 2 and Tables I and II. For these baryons the
theoretical estimates are much lower than the experimental values, more or
less independent of the parameter set employed in the calculation. 
This comes to no surprise since the mass splitting between the octet and the
decuplet is usually attributed to the spin--spin interaction among quarks,
which we have till now neglected.

The latter can be identified with the color 
magnetic interaction, which is largely responsible for the octet--decuplet 
mass splitting. In the spirit of Ref.~\cite{Hatsuda94,Kuni90,Hatsuda92}, 
we shall consider the spin interaction as a weak perturbation between 
constituent quarks, by assuming the NJL model as a field theoretical 
version of the constituent quark model itself.

The usual spin--spin interaction derived from one gluon exchange refers 
to non--relativistic quarks with constituent masses. It 
is generally written as\cite{Hatsuda94}:
\be
V_{spin}=a\sum_{i<j}\frac{{\bf\sigma}_i{\bf\sigma}_j}{M_iM_j},
\label{Vspin}
\ee
where  M$_i$ and M$_j$ are the (constant) masses in the vacuum. Here we 
do not take into account the density dependence of the quark masses 
inside the baryon since the coefficients in the mass formula which 
reproduce the Gell Mann Okubo relation include infinite higher orders 
in the current masses $m_i$ hidden in $M_i$. As a safe compromise we 
adopt the zero--range spin interaction (\ref{Vspin}), yet accounting for
the density distribution of quarks in the baryon. Hence we write:
\be
\Delta M_{spin}=a\sum_{i<j}\frac{1}{M_iM_j}\int 
d\vec{r}_1d\vec{r}_2\rho_i\left(\vec{r}_1\right)
\rho_j\left(\vec{r}_2\right)<{\bf\sigma}_i{\bf\sigma}_j>
\delta\left(\vec{r}_1-\vec{r}_2\right),
\ee
and, by performing one of the integrations, we finally find:
\be
\Delta M_{spin}=a\sum_{i<j}\frac{1}{M_iM_j}\int d\vec{r}\rho_i
\left(\vec{r}\right)
\rho_j\left(\vec{r}\right)<{\bf\sigma}_i{\bf\sigma}_j>.
\ee

\begin{figure}[htb]
\begin{center}
{\epsfig{file=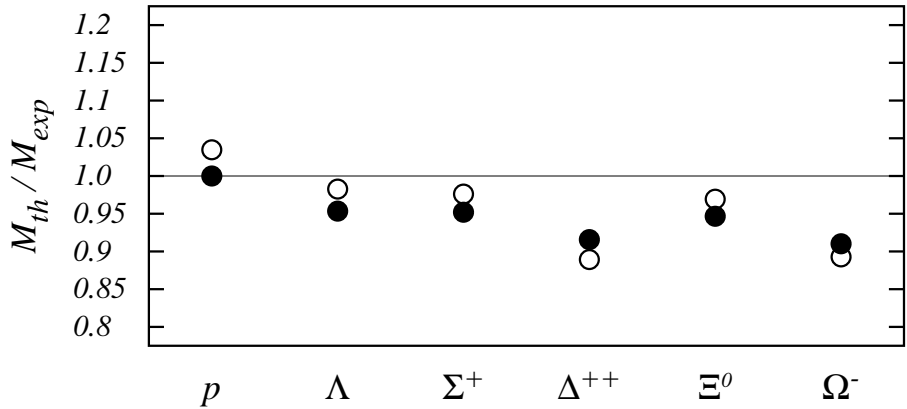,width=.85\textwidth}}
\caption{Theoretical-experimental baryon mass ratio as obtained with the
parameter set 1: empty circles correspond to the calculation without spin 
(see Fig.~\protect\ref{fig2}), full circles include spin corrections.}
\label{fig3}
\end{center}
\end{figure}

We found it convenient to fix the $a$ parameter in order to reproduce the 
nucleon mass (rather than the $N-\Delta$ mass splitting) thus 
obtaining, for the calculation with the parameters of set 1, 
$a=0.72$. The resulting masses are reported in the last row
of Table~II, while their ratios to the experimental masses are shown in
 in Fig.~\ref{fig3} (full circles). One can see that, by including the 
spin correction, the octet masses are not much modified,
 while the $\Delta^{++}$ and $\Omega^-$ masses, though somewhat improved, 
remain smaller than the experimental values, the discrepancy being limited 
within $9\%$.
Obviously a larger value for the $a$ parameter would improve the decuplet
masses, but worsening, at the same time, the octet masses.
Using the parameter set 2 the situation cannot be sensibly improved by
the introduction of spin corrections, since all values obtained for the baryon 
masses (including the proton one) are already somewhat smaller than the 
corresponding experimental masses. 

\section{Conclusions} 

We have considered the three--flavour effective Lagrangian of the NJL 
model in order to determine, within the mean field approximation, the 
 effective vacuum pressure acting on a nucleon; this allowed us to fix
in an unambiguous way the parameter of the quark wave functions in a 
baryon, which were heuristically assumed to be bound state wave functions
of Gaussian type. Furthermore we have evaluated the masses of the baryon 
octet by implementing the local density approximation on the quark 
energies obtained in a uniform and isotropic system.

The whole approach relies on the usual interpretation that the NJL 
dynamical masses, related to the quark condensate, provide a consistent
connection with the constituent quark model. However, at variance with
previous works, we take into account the density dependence of the 
constituent masses, a feature which appears to be crucial in order to 
obtain a sensible determination of the various baryonic masses.

 We have calculated the nucleon radius under the assumption that only 
the vacuum pressure is responsible for the baryon stability. 
The results we obtained are fairly good: in particular 
the baryonic masses turn out to be close to the experimental values,
already when we limit ourselves to take  into account the quark 
dynamics of the NJL model. Spin corrections are shown to slightly improve the
results for the decuplet masses, though they remain smaller than the 
experiment.

We found a remarkable stability of our results with respect to 
different assumptions for the quark density in the baryon; yet one 
important ingredient of our calculation is the density dependence of
the dynamical masses, which provides, by itself, a realistic 
determination of the octet baryonic masses, even without resorting to
spin and/or confining corrections, as usually adopted in the context
of the constituent quark model.

{\bf Acknowledgments}\\
We are deeply indebted with Dr. Thomas Gutsche for many valuable
comments and for a careful reading of the manuscript. We also thank
Prof. A. Gal for very stimulating remarks.

\end{document}